\documentclass[%
 reprint,
 amsmath,amssymb,
 aps,
]{revtex4-2}

\usepackage{graphicx}
\usepackage{dcolumn}
\usepackage{bm}
\usepackage{hyperref}
\usepackage[english]{babel}
\usepackage{xcolor}

\newcommand{\e}{\varepsilon}
\newcommand{\w}{\omega}
\newcommand{\kp}{\kappa}

\newcommand{\ee}{\mathrm{e}}
\newcommand{\ii}{\mathrm{i}}
\newcommand{\dd}{\,\mathrm{d}}

\newcommand{\mF}{\mathcal{F}}
\newcommand{\mO}{\mathcal{O}}
\newcommand{\mQ}{\mathcal{Q}}
\newcommand{\mR}{\mathcal{R}}
\newcommand{\mS}{\mathcal{S}}

\newcommand{\bL}{{\bf L}}

\begin{document}

\preprint{APS/123-QED}

\title{
Phase reduction explains chimera shape: when multi-body interaction matters
}

\author{Erik T. K. Mau}
\email[]{erikmau@uni-potsdam.de}
\affiliation{Department of Physics and Astronomy, University of Potsdam, 
Karl-Liebknecht-Str. 24/25, D-14476 Potsdam-Golm, Germany}
 
\author{Oleh E. Omel’chenko}
\email{omelchenko@uni-potsdam.de}
\affiliation{Department of Physics and Astronomy, University of Potsdam, 
Karl-Liebknecht-Str. 24/25, D-14476 Potsdam-Golm, Germany}

\author{Michael Rosenblum}
\email[]{mros@uni-potsdam.de}
\affiliation{Department of Physics and Astronomy, University of Potsdam, 
Karl-Liebknecht-Str. 24/25, D-14476 Potsdam-Golm, Germany}

\date{\today}

\begin{abstract}
We present an extension of the Kuramoto-Sakaguchi model for networks, deriving the second-order phase approximation for a paradigmatic model of oscillatory networks - an ensemble of non-identical Stuart-Landau oscillators coupled pairwisely via an arbitrary coupling matrix. We explicitly demonstrate how this matrix translates into the coupling structure in the phase equations. To illustrate the power of our approach and the crucial importance of high-order phase reduction, we tackle a trendy setup of non-locally coupled oscillators exhibiting a chimera state. We reveal that our second-order phase model reproduces the dependence of the chimera shape on the coupling strength that is not captured by the typically used first-order Kuramoto-like model. Our derivation contributes to a better understanding of complex networks' dynamics, establishing a relation between the coupling matrix and multi-body interaction terms in the high-order phase model. 
\end{abstract}


\maketitle

\section{Introduction}

A central problem of network science is an account of multi-body interactions~\cite{battiston2020,*battiston2021,*battiston2022,*boccaletti2023},
which can be responsible for various dynamical phenomena~\cite{skardal2020,*bick2023c}. In oscillatory networks, such interactions naturally appear on the level of phase description via the terms depending on the phases of three or more oscillators. In particular, such terms emerge due to the high-order phase reduction of pairwisely coupled oscillators. In this Letter, we elaborate on this scenario and treat a system of $N$ Stuart-Landau (SL) oscillators:
\begin{align}
    \dot z_i = (\eta_i +  \ii \nu_i)z_i - \eta_i(1 +\ii \beta_i)|z_i|^2 z_i
    + \e  \sum_{j=1}^N L_{ij} \ee^{\ii \alpha_{ij}} z_j\,,
    \label{SL_original}
\end{align}
where $z_i$ are complex variables and $\eta_i,\nu_i,\beta_i \in \mathbb{R}$,
$\eta_i > 0$, are 
parameters~\footnote{Notice that the form of Eq.~(\ref{SL_original})
ensures that all SL oscillators have limit cycles with $|z_i| = 1$.
But this does not restrict the generality of our approach.
Indeed, we can start with the general form of the SL equation,
writing the nonlinear term as $-(\zeta_i+\ii\beta_i)$ with $\zeta_i > 0$.
Then, we can transform this equation to Eq.~(\ref{SL_original})
by substituting $z_i\mapsto z_i\sqrt{\eta_i/\zeta_i}$, $\beta_i\mapsto \beta_i \zeta_i$,
and rescaling the matrix $\bL$.}.
We emphasize that oscillators are non-identical and the real-valued coupling matrix $\bL$ and phase shifts $\alpha_{ij}$ are arbitrary. Parameter $\e$ explicitly quantifies the coupling strength; in the following, we assume that it is small compared to $\eta_i$, which quantify the stability of the limit cycles. The choice of the model is motivated by the well-known fact that the phase approximation of SL systems in the first order in $\e$ yields the celebrated Kuramoto-Sakaguchi (KS) model~\cite{kuramoto1975,*kuramoto1984,*sakaguchi1986,*nakao2016,*kuramoto2019,*pietras2019}.
Here, we derive the second-order phase approximation for system~(\ref{SL_original}), providing an extension of the KS system on networks for coupling that is not too weak:
\begin{align}
    \dot{\phi}_i = \w_i + \e \mR_i(\Vec{\phi}-\phi_i \Vec{e}) + \e^2 \mS_i(\Vec{\phi},\phi_i)
    \,.
    \label{phmodel1}
\end{align}
Here, $\phi_i$ is the phase of the $i$th oscillator, $\Vec{\phi}=(\phi_1,\phi_2,\ldots,\phi_N)^\top$, 
$\Vec{e} = (1, \dots, 1)^\top$, and first- and second-order coupling functions $\mR_i$ and $\mS_i$ are explicitly given below 
by Eqs.~(\ref{eq:Qi_0},\ref{eq:Qi_1_result}).  
The crucial feature of Eq.~(\ref{phmodel1}) is that the KS terms  $\mR_i$ depend only on phase differences while $\mS_i$ contains both difference and triplet terms.

\section{Main Results}

\begin{figure}
\includegraphics[height=0.32\columnwidth]{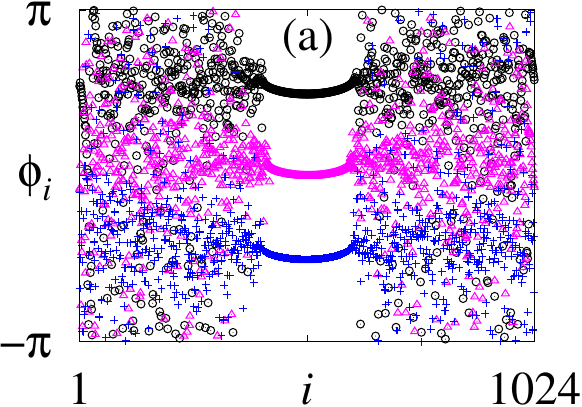}\hspace{0.03\columnwidth}
\includegraphics[height=0.32\columnwidth]{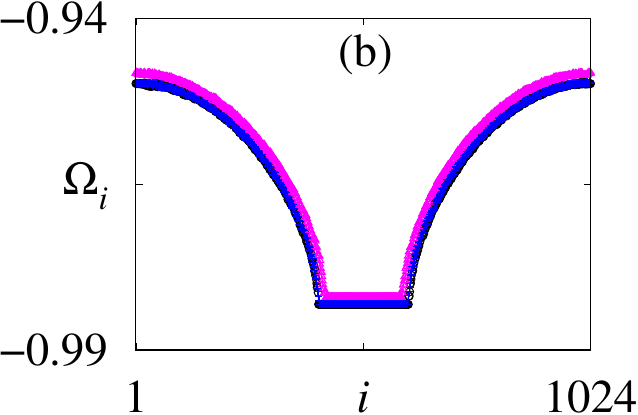}\\[5mm]
\includegraphics[height=0.32\columnwidth]{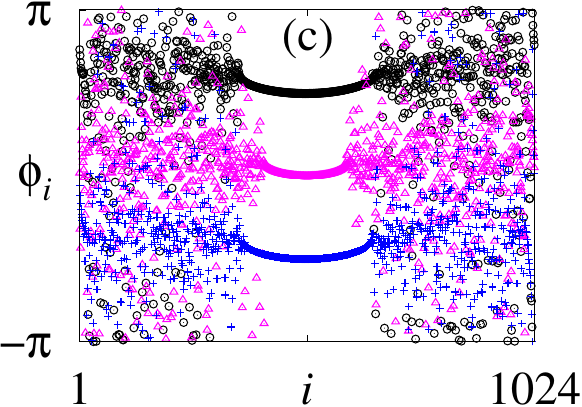}\hspace{0.03\columnwidth}
\includegraphics[height=0.32\columnwidth]{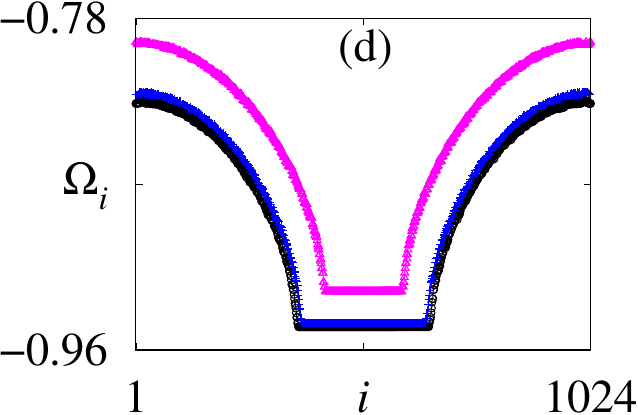}\\[5mm]
\includegraphics[width=0.99\columnwidth]{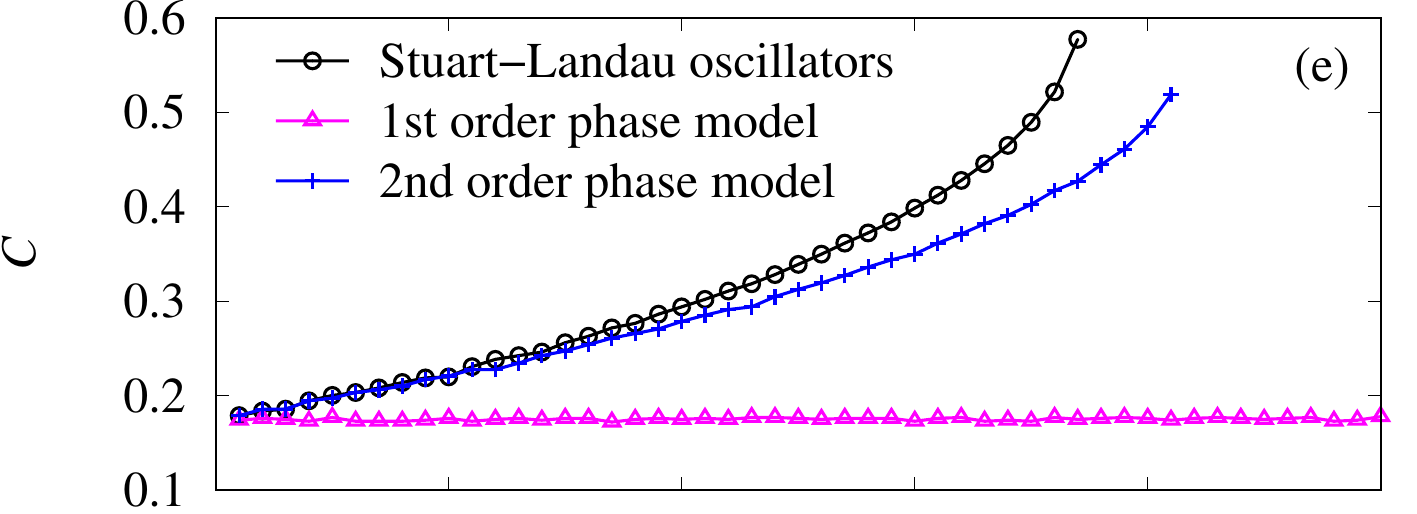}\\[2mm]
\includegraphics[width=0.99\columnwidth]{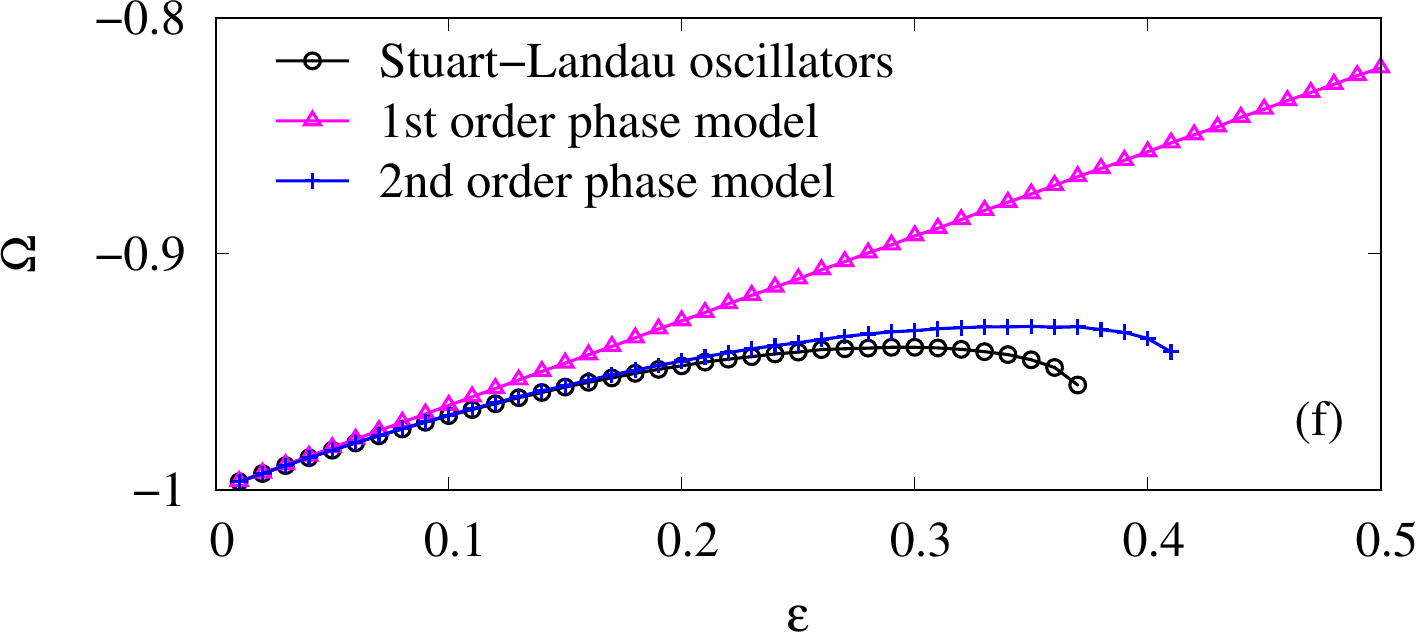}
\caption{Chimera states in the system of $N=1024$ nonlocally coupled identical 
Stuart-Landau oscillators (black circles)
and in the corresponding phase-reduced models
of the first (magenta triangles) and second (blue crosses) orders. Phase snapshots $\phi_i$
and time-averaged phase velocities $\Omega_i$
are shown for coupling strengths
$\e = 0.05$ (a,b) and $\e = 0.2$ (c,d).
Panel (e) exhibits the dependence of the relative size $C$
of the chimera's coherent region (i.e., the fraction of frequency-locked oscillators) on the coupling strength $\e$.
The time-averaged phase velocity of coherent oscillators $\Omega$ versus the coupling strength $\e$ is shown in panel (f).
Numerical results in this figure demonstrate that the second-order phase approximation successfully reproduces the shape and frequency of the coherent domain for chimera states with quite a strong coupling $\e\lessapprox 0.25$.
Like the SL system, the 2nd-order phase model shows the disappearance of the chimera for $\e\approx 0.4$.
}
\label{Fig:MainResults}
\end{figure}

The high-order terms are important because (i) they increase the phase-reduced model's accuracy, and (ii) they can describe the effects of the original dynamics not captured by the first-order approximation.
To illustrate this, we consider a particular
but important setup - a ring of
identical nonlocally coupled SL oscillators
- exhibiting symmetry-breaking solutions known
as chimera states~\cite{kuramoto2002,abrams2004,*panaggio2015,*schoell2016,*kemeth2016,*omelchenko2018,*parastesh2021}. 
In this case, $\w_i=\w, \mR_i=\mR, \mS_i=\mS$ do not depend on the index $i$, so Eq.~(\ref{phmodel1}) simplifies to 
\begin{align}
    \dot{\phi}_i = \w + \e \mR(\Vec{\phi}-\phi_i \Vec{e}) + \e^2 \mS(\Vec{\phi},\phi_i)\;,
\label{phmodel2}
\end{align}
where the exact relation between the matrix $\bL$ and coupling functions $\mR,\mS$ is given by Eqs.~(\ref{eq:L_kernel},\ref{chimera_2nd_order}) below.
Recall that chimera state is a dynamical pattern
with self-organized domains of synchronized (coherent)
and desynchronized (incoherent) oscillators.
We emphasize that in the first-order approximation, the spatial distribution of coherent and incoherent oscillators on the ring is independent of coupling strength. Indeed, if one neglects the terms $\sim \e^2$, goes to the co-rotating frame, and then rescales the time, $t\mapsto t/\e$, then the resulting equation does not depend on $\e$. We demonstrate that, however, the full SL model exhibits dependence of the chimera shape on $\e$, and so does the second-order phase model~(\ref{phmodel2}).
For this, we revisit the classical model where chimera states were discovered by Kuramoto and Battogtokh~\cite{kuramoto2002}.
We choose the nonlocal coupling in the form
\begin{equation}
L_{ij} = \frac{2}{N} G_\mathrm{exp}\left( \frac{2 d_{ij}}{N} \right) - \delta_{ij}\;,
\label{Coupling:Exp}
\end{equation}
where $\delta_{ij}$ is the Kronecker delta,
\begin{equation}
G_\mathrm{exp}(x) = e^{-2x}/(1 -e^{-2}) 
\label{G_exp}
\end{equation}
is an exponentially decaying function
associated with the adiabatic elimination
of the diffusive mediator in a two-component system,
and
$
d_{ij} = \min\{ |i - j|, N - |i - j| \}
$
is the distance between the $i$th and $j$th oscillators
on the ring of length $N$.
(Note the normalization factors in~\eqref{Coupling:Exp} and~\eqref{G_exp} guarantee that the interaction is asymptotically balanced, i.e., $\sum_{j=1}^N L_{ij} = 0$ for $N\to\infty$.)
As for other parameters, we adopt them from~\cite{sethia2013a}.
Namely, we fix the parameters of SL oscillators
as $\eta_i = \beta_i = 1$ and $\nu_i = 0$, and we choose
$\alpha_{ij} = \alpha = -\arctan(0.9)$ for all $i$ and $j$.
Next, we compare the results of the numerical simulation for 
the SL ring with those for the first- and second-order phase 
models~\footnote{To calculate the solution
of the system~(\ref{SL_original}), we used the standard
fourth-order Runge-Kutta integrator
with a fixed time step $dt = 0.002/\e$.
We discarded transients of length $2000/\e$ time units
and used the next $500/\e$ time units
to calculate the time-averaged phase velocities $\Omega_i$. 
The chimera state for $\e = 0.01$ was obtained
using the coherent-incoherent initial condition:
$z_i = 1$ for $i < 0.2 N$,
and $z_i = e^{i \xi_i}$ otherwise,
where $\xi_i$ is a random variable
uniformly distributed in the interval $(-\pi,\pi]$.
The parameter sweep in Figure~\ref{Fig:MainResults}
was calculated by adiabatically increasing $\e$
with a step of $d\e = 0.01$.
A similar numerical protocol was also used
to calculate solutions of phase-reduced models.}.
The computations for the SL model clearly demonstrate
that the shape of the chimera state
changes for different values of $\e$
even if $\e$ remains relatively small,
see Fig.~\ref{Fig:MainResults}.
Moreover, the same simulations show
that as $\e$ increases, the chimera states
cease to exist at a fold bifurcation,
while the first-order phase model does not signal this.
However, if, along with the $\e^1$-terms,
we also keep the $\e^2$-terms in Eq.~(\ref{phmodel2}),
the performance of the phase reduction improves significantly.
In particular, Fig.~\ref{Fig:MainResults}
shows that the second-order phase model adequately 
reproduces the changes in the shape
of the chimera state for the variation of $\e$,
which the first-order phase model cannot capture.

\section{Second-order phase reduction for the SL network}
Our derivation of Eq.~\eqref{phmodel1} follows~\cite{gengel2021,mau2023c} and consists of
two main steps. The first step is the (invertible) transformation of each oscillator to its phase-isostable normal form~\cite{shilnikov1998, wilson2018a}, expressing its state by phase $\phi_i \in [0,2\pi)$ and isostable amplitude $r_i\in \mathbb{R}$ that is zero at the limit cycle.
The actual phase reduction relies
on the weak coupling assumption.
In this case, our coupled SL system with $\eta_i > 0$
has an attractive smooth invariant $N$-dimensional torus~\cite{fenichel71}
parameterized by the phases $\phi_i$.
Thus, focusing on the system's long-term behavior, we take all isostable amplitudes as functions of the phases, i.e., $r_i = R_i(\Vec{\phi},\e)$.
In the second step, we suppose
that each $R_i(\Vec{\phi},\e)$ is smooth
and $R_i(\Vec{\phi},0)=0$ so that
\begin{align}
R_i(\Vec{\phi},\e) = R_{i;1}(\Vec{\phi})\e + \mO(\e^2)
\label{eq:R_expansion}
\,.
\end{align}
Then, we use ansatz \eqref{eq:R_expansion}
to obtain the first and second-order terms in \eqref{phmodel1} from the formal asymptotic expansion.

For convenience, we first switch to physically meaningful parameters for each unit: frequency of the limit-cycle oscillation, $\w_i = \nu_i - \eta_i \beta_i$, Floquet exponent of the cycle, $\kappa_i = -2\eta_i<0$,
and the non-isochronicity parameter $\gamma_i = \arctan(\beta_i)$.
With $c_i=\kp_i\tan(\gamma_i)/2$, Eq.~\eqref{SL_original} becomes
\begin{align*}
    \dot{z}_i
    =& \left( i\w_i + (\frac{\kp_i}{2}+\ii c_i)(|z_i|^2-1) \right)z_i 
    + \e  \sum_{j=1}^N L_{ij}\ee^{\ii \alpha_{ij}} z_j
    \,.
\end{align*}
Next, we re-write the system in polar coordinates $\rho_i=|z_i|$ and $\theta_i=\text{arg}(z_i)$: 
\begin{align}
    \dot{\theta}_i &= \w_i + c_i (\rho_i^2-1) 
    +\e \sum_{j=1}^N \frac{L_{ij}\rho_j}{\rho_i} \sin(\theta_j-\theta_i + \alpha_{ij})\;, 
    \label{eq:model_polar_theta} \\
    \dot{\rho}_i &= \frac{\kp_i}{2}\rho_i(\rho_i^2-1)
    +\e \sum_{j=1}^N L_{ij} \rho_j \cos(\theta_j-\theta_i + \alpha_{ij})
    \label{eq:model_polar_rho}
    \,.
\end{align}
We recall that without loss of generality, we wrote Eq.~(\ref{SL_original}) so that on the limit cycle $\rho_i=1$ for all $i$.

It is well-known~\cite{wilson2018a}
that for each SL oscillator of the form
$\dot{\theta} = \omega + c (\rho^2 - 1)$,
$\dot{\rho} = \kappa \rho (\rho^2 - 1) / 2$,
there exists a phase-isostable transformation:
\begin{align}
    \phi &= \Phi(\theta, \rho) = \theta - \tan(\gamma) \ln(\rho) \,,\quad 
    \label{eq:phi_transform}
     \\
    r &= P(\rho) = (\rho^2-1)/(2\rho^2) 
    \label{eq:r_transform}\,
\end{align}
with $\tan(\gamma) = 2 c / \kappa$,
which is defined by two characteristic properties. First, in the new coordinates,
the dynamics of the amplitude $r$ and the phase $\phi$
decouple and take on a simple linear form
$\dot{\phi}=\w$ and $\dot{r} = \kp r$.
Second, on the limit cycle ($\rho=1$), it holds $r=0$ and $\phi=\theta$. Note that the transformation
(\ref{eq:phi_transform},\ref{eq:r_transform})
defined in this way~\footnote{The phase-isostable transformation   (\ref{eq:phi_transform},\ref{eq:r_transform})
is obtained by solving two differential equations $\dot\phi=\w=\frac{d}{dt}\Phi(\theta,\rho)$ and $\dot r=\kappa r=\frac{d}{dt}P(\rho)$, which explicitly read 
$\omega = (\omega + c (\rho^2 - 1)) \partial_\theta \Phi + 0.5 \kappa \rho (\rho^2 - 1) \partial_\rho \Phi$ and
$\kappa P = 0.5 \kappa \rho (\rho^2 - 1) \partial_{\rho} P$, 
with two boundary conditions
$\Phi(\theta,1) = \theta$ and $P(1) = 0$.
Note that although Eq.~(\ref{eq:phi_transform})
is a unique solution,
Eq.~(\ref{eq:r_transform}) is defined
only up to a scaling factor.}
is isomorphic
on the entire basin of attraction
of the limit cycle, i.e. for all $\rho >0$~\cite{shilnikov1998}.
In addition, the isochrones of the SL oscillator $\Phi(\theta, \rho) = \mathrm{const}$ 
are logarithmic spirals with the inclination angle $\gamma$
relative to the outward normal on the limit cycle ($\rho = 1$).
For $\gamma = 0$, these spirals turn into straight rays
and Eq.~\eqref{eq:phi_transform}
degenerates into the identity $\phi = \theta$.

Using~(\ref{eq:phi_transform},\ref{eq:r_transform})
and the identities
$\dot{\phi}_i = \partial_{\theta_i} \Phi_i\cdot\dot{\theta}_i + \partial_{\rho_i} \Phi_i \cdot\dot{\rho}_i$ and $\dot{r}_i = \partial_{\rho_i} P_i \cdot \dot{\rho_i}$,
we write system~(\ref{eq:model_polar_theta},\ref{eq:model_polar_rho}) in the form:
\begin{align}
    \dot{\phi}_i = \w_i + \e \mQ_i(\Vec{\phi}, \Vec{r})\;, \qquad
    \dot{r}_i = \kp_i r_i + \e \mF_i(\Vec{\phi}, \Vec{r})
    \label{eq:phase_iso_dynamic}
    \,,
\end{align}
where $\mQ_i$ is the phase coupling function
\begin{align}
    \mQ_i(\Vec{\phi}, \Vec{r})
    &= \sum_{k=1}^N L_{ik} \frac{\rho_k \sin(\theta_k - \theta_i - \gamma_i + \alpha_{ik})}{\rho_i \cos(\gamma_i)}
    \label{eq:mQ_i}
\end{align}
and $\mF_i$ is the amplitude coupling function
\begin{align}
    \mF_i(\Vec{\phi}, \Vec{r})
    &= \sum_{k=1}^N L_{ik}
    \frac{\rho_k \cos(\theta_k - \theta_i + \alpha_{ik})}{\rho_i^3}
    \label{eq:mF_i}
    \,.
\end{align}
In Eqs.~(\ref{eq:mQ_i},\ref{eq:mF_i}), the polar coordinates $\rho_i, \theta_i$ are functions of $\phi_i, r_i$ according to transformations~(\ref{eq:phi_transform},\ref{eq:r_transform}). In the following, for brevity, we omit the arguments $\Vec{\phi}$ and $\Vec{r}$.

The first-order term $\mR_i$ in model~\eqref{phmodel1} follows from the phase coupling function evaluated at the limit cycle, i.e., $\mR_i = \left. \mQ_i \right|_{\Vec{r}=0}$.
Thus, from Eq.~\eqref{eq:mQ_i} we obtain
\begin{align}
    \mR_i = \sum_{j=1}^N \frac{L_{ij}}{\cos(\gamma_i)} \sin(\phi_j - \phi_i - \gamma_i + \alpha_{ij})
     \label{eq:Qi_0}
     \,,
\end{align}
which is the well-known Kuramoto-Sakaguchi~\cite{sakaguchi1986} term.

For the second-order term $\mS_i$,
we substitute formula~\eqref{eq:R_expansion} into
the first equation of system~\eqref{eq:phase_iso_dynamic}
and expand $\mQ_{i}$ in powers of $\e$. This yields
$\mS_i = \sum_{j=1}^N \left.\partial_{r_j} \mQ_{i} \right|_{\Vec{r}=0} R_{j;1}$.
On the other hand, by substituting formula~\eqref{eq:R_expansion} into the second equation of system~\eqref{eq:phase_iso_dynamic} we find that $R_{j;1}$ is determined by the linear partial differential equation (PDE)
\begin{align}
    \kp_j R_{j;1} - \sum_{k=1}^N \w_k \partial_{\phi_k} R_{j;1} = -\left. \mF_{j} \right|_{\Vec{r}=0}
    \label{eq:PDE}
    \,,
\end{align}
which can be solved using the Fourier transform \cite{mau2023c}. Thus we obtain
$R_{j;1} = -\Xi_j[\left. \mF_{j} \right|_{\Vec{r}=0}]$,
where
\begin{align}
    \Xi_j[f] 
    &= \frac{1}{(2\pi)^N} \int_0^{2\pi} f(\Vec{\phi}-\Vec{s}) \sum_{\Vec{m} \in \mathbb{Z}} \frac{\ee^{-\ii \Vec{m} \cdot \Vec{s}}}{\kp_j + \ii \Vec{m}\cdot \Vec{\w}}
    \dd \Vec{s}
    \label{eq:Xi_def}
\end{align}
denotes the solution operator of PDE~\eqref{eq:PDE}
with arbitrary r.h.s. $f$, and $\Vec{m} \cdot \Vec{s}$ and $\Vec{m} \cdot \Vec{\w}$ are scalar products.
In conclusion, as a result of this consideration, we obtain
\begin{align}
    \mS_i 
    &= - \sum_{j=1}^N 
    \left.\partial_{r_j} \mQ_{i} \right|_{\Vec{r}=0} 
    \cdot \Xi_j[\left. \mF_{j} \right|_{\Vec{r}=0}]
    \label{eq:Qi_1}
    \,,
\end{align}
where all terms in the resulting formula
can be found explicitly.
Indeed, using (\ref{eq:phi_transform},\ref{eq:r_transform},\ref{eq:mQ_i})
and the identity $\left. \partial_{r_j} \rho_i \right|_{\Vec{r}=0} = \delta_{ij}$, we calculate
\begin{align}
    \left.\partial_{r_j} \mQ_{i} \right|_{\Vec{r}=0}
    &= \sum_{k=1}^N L_{ik}\left. \partial_{r_j}  q_{ik} \right|_{\Vec{r}=0}
    \label{eq:Qi_ej}
\end{align}
with
\begin{align}
    \left. \partial_{r_j}  q_{ik} \right|_{\Vec{r}=0}
    &= \frac{(\delta_{kj} - \delta_{ij})\sin(\phi_k - \phi_i + \gamma_j - \gamma_i + \alpha_{ik})}{\cos(\gamma_i)\cos(\gamma_j)}
    \label{eq:q_ik_deriv_0}
    \,.
\end{align}
Next, substituting $\Vec{r}=0$ into Eq.~\eqref{eq:mF_i} we obtain
\begin{align*}    
    \left. \mF_j \right|_{\Vec{r}=0} = \sum_{k=1}^N L_{jk} \cos(\phi_k - \phi_j + \alpha_{jk})
    \,.
\end{align*}
Thus, using Eq.~\eqref{eq:Xi_def} we find:
\begin{align}    
    \Xi_j[\left. \mF_j \right|_{\Vec{r}=0}] 
    = \sum_{k=1}^N  \frac{L_{jk}}{\kp_j} \cos(\Delta_{jk}) \cos(\phi_k-\phi_j-\Delta_{jk} + \alpha_{jk})
    \label{eq:Xi_i_F_i}
\end{align}
where $\Delta_{jk} = \arctan((\w_j-\w_k)/\kp_j)$
is the phase lag due to the frequency mismatch.
Finally, substituting Eqs.~(\ref{eq:Qi_ej},\ref{eq:q_ik_deriv_0},\ref{eq:Xi_i_F_i}) into Eq.~\eqref{eq:Qi_1} and using basic trigonometric identities, we obtain
\begin{align}
    \mS_i 
    &= \frac{1}{2} \sum_{j=1}^N \sum_{k=1}^N \left( \hat{S}_{ijk} - \Tilde{S}_{ijk} \right)
    \label{eq:Qi_1_result}
\end{align}
with 
\begin{align}
    \hat{S}_{ijk} 
    &=
    \frac{L_{ij}L_{ik} \cos(\Delta_{ik})}{\kp_i \cos^2(\gamma_i)}\biggl( 
    \nonumber \\
    &\sin(\phi_j-\phi_k + \Delta_{ik} + \alpha_{ij} - \alpha_{ik})
    \nonumber \\
    \qquad + &\sin(\phi_j + \phi_k - 2\phi_i - \Delta_{ik} + \alpha_{ij}+\alpha_{ik}) \biggr) 
    \label{eq:S_ijk}
\end{align}
and 
\begin{align}
    \Tilde{S}_{ijk} 
    &= 
    \frac{L_{ij}L_{jk} \cos(\Delta_{jk})}{\kp_j \cos(\gamma_i) \cos(\gamma_j)}\biggl( 
    \nonumber \\
    &\sin(2\phi_j - \phi_k - \phi_i + \gamma_j-\gamma_i + \Delta_{jk} + \alpha_{ij} - \alpha_{jk})
    \nonumber \\
    \qquad + &\sin(\phi_k - \phi_i + \gamma_j - \gamma_i - \Delta_{jk} + \alpha_{ij} + \alpha_{jk})
    \biggr)
    \,.
    \label{eq:S_ijk_}
\end{align}
Equations~(\ref{eq:Qi_0},\ref{eq:Qi_1_result}-\ref{eq:S_ijk_}) complete the second-order phase model~\eqref{phmodel1} for heterogeneous SL oscillators coupled pairwisely via an arbitrary complex-valued matrix.
We emphasize some facts about them.
$\mS_i$ is a sum of two pairwise and two triplet (multi-body interaction) terms. Structurally,  all terms can be divided into direct and mediated interactions, indicated by $\hat{S}_{ijk}$ and $\Tilde{S}_{ijk}$, respectively.
The pairwise and triplet terms in $\hat{S}_{ijk}$ share the coefficient $L_{ij} L_{ik}$, so they influence node $i$ only if both nodes $j$ and $k$ have a structural connection to $i$. Interestingly, for identical oscillators, $\w_i = \w$, the pairwise terms in $\hat{S}_{ijk}$ cancel when summed over $j$ and $k$.
On the contrary, the terms in $\Tilde{S}_{ijk}$ share the coefficient $L_{ij}L_{jk}$, thus they include the pairwise and triplet interactions of node $i$ with all nodes $k$ that are connected via a mediator $j$.~\footnote{A similar derivation for a slightly different system has been done in~\cite{bick2024a}, where the authors used the terminology of ``virtual pairwise connections'' for the pairwise term in $\Tilde{S}_{ijk}$ because of the mediated coupling, that does not need a structural connection between oscillators $i$ and $k$ to facilitate a functional coupling.}
Finally, we note that by definition $\cos(\gamma_i)>0$ and $\cos(\Delta_{ij})>0$  for all indices.

\subsection{A particular case: global coupling}
For global diffusive coupling we have $L_{ij} = \frac{1}{N} - \delta_{ij}$. For brevity, we consider isochronous oscillators ($\gamma_i = 0$) with identical Floquet exponents $\kp_i=\kp$ and phase lags $\alpha_{ij} = \alpha$ but non-identical frequencies. Then, Eqs.~(\ref{eq:Qi_0},\ref{eq:Qi_1_result}-\ref{eq:S_ijk_}) yield
\begin{align}
    \dot{\phi}_i =& ~\w_i -\e \sin\alpha +
     \frac{\e}{N} \sum_{j=1}^N \sin(\phi_j - \phi_i + \alpha) 
    \nonumber \\
    +& \frac{\e^2}{2\kp N^2} \sum_{j,k=1}^N
    \biggl(
    \cos(\Delta_{ik})\sin(\phi_j-\phi_k + \Delta_{ik})
    \nonumber \\
    +&\cos(\Delta_{ik})\sin(\phi_k+\phi_j-2\phi_i -\Delta_{ik} + 2\alpha)
    \nonumber \\ 
    -&\cos(\Delta_{jk})\sin(2\phi_j-\phi_k-\phi_i + \Delta_{jk})
    \nonumber \\
    -&\cos(\Delta_{jk})\sin(\phi_k-\phi_i -\Delta_{jk} +2\alpha)
    \biggr)
    \,.
    \label{eq:model_global_coupling}
\end{align}
Note that the frequency differences, encoded in $\Delta_{ij} = \arctan((\w_i-\w_j)/\kp)$, influence both the amplitude and the phase shift of the second-order coupling terms. For identical frequencies, $\Delta_{ij} =0$, our model reduces to that of  León and Pazó (up to a choice of parameters)~\footnote{Using a different approach, León and Pazó derived second and third-order phase models for globally coupled identical SL oscillators~\cite{leon2019}}.

\subsection{A particular case: non-locally coupled identical units}
Now, we use formulas (\ref{eq:Qi_0},\ref{eq:Qi_1_result}-\ref{eq:S_ijk_})
to write an explicit form of the second-order phase approximation
for a ring of identical SL oscillators, i.e., for the chimera setup.
We take $\w_i=\w$, $\gamma_i=\gamma$, and $\kp_i=\kp$~\footnote{We remind that these parameters
are related to the parameters $\eta_i$, $\beta_i$ and $\nu_i$
in Eq.~(\ref{SL_original}) by formulas
$\omega_i = \nu_i - \eta_i \beta_i$, $\kappa_i = - 2\eta_i$ and $\gamma_i = \arctan(\beta_i)$.}.
Then, $\Delta_{ij} = 0$. 
The non-local coupling is organized in the standard way:
\begin{align}
    L_{ij} = \frac{1}{N} G(i-j) - \delta_{ij}\;,
    \label{eq:L_kernel}
\end{align}
where $G$ is a periodic function $G(i) = G(i\pm N)$, which can be symmetric or not, and we also fix $\alpha_{ij}=\alpha$.
Inserting Eq.~\eqref{eq:L_kernel} into expressions for 
$\mR_i$ and $\mS_i$ we obtain
\begin{align}
    \dot{\phi}_i = \Bar{\w} + & 
    \frac{\Bar{\e}}{N} \sum_{j=1}^N G(i-j) \sin(\phi_j - \phi_i - \gamma + \alpha) 
    \nonumber \\
    ~+ \frac{\Bar{\e}^2}{2\kp N^2} & \sum_{j,k=1}^N
    \biggl(
    G(i-j)G(i-k) \sin(\phi_j+\phi_k-2\phi_i + 2\alpha)
    \nonumber \\ 
     & -G(i-j)G(j-k) \sin(2\phi_j-\phi_k-\phi_i)
    \nonumber \\
     & -G(i-j)G(j-k) \sin(\phi_k-\phi_i +2\alpha)
    \biggr) \,,
    \label{chimera_2nd_order}
\end{align}
where $\Bar{\e} = \e/\cos(\gamma)$ and $\Bar{\w} = \w + \Bar{\e} \sin(\gamma-\alpha)$,
cf. Eq.~(\ref{phmodel2}). We used Eq.~(\ref{chimera_2nd_order})
for numerical simulations shown in Fig.~\ref{Fig:MainResults} for a special choice of $G$, see Eq.~\eqref{G_exp}; this simulation 
validates the obtained approximation for moderate values of $\e$.
Notice that Eq.~(\ref{chimera_2nd_order}) remains phase-shift invariant,
therefore going to a corotating frame and rescaling time
we can show that the resulting dynamics depend on the ratio $\e/\kappa$,
but not on $\e$ and $\kappa$ separately.

\section{Discussion}
We summarize our results. We presented second-order phase reduction
for a network of non-identical and arbitrarily coupled SL oscillators. The main effect is that pairwise coupling of the SL units results in the emergence of hypernetwork with triplet interactions on the level of phase description. Our derivation yields the correct structure of multi-body interaction in the physically motivated extension of the Kuramoto-Sakaguchi model on networks, in particular, for globally coupled units, see Eq.~\eqref{eq:model_global_coupling}.
Furthermore, we provide the phase equation describing chimera states with an account of second-order terms. The relation of the coupling strength and the Floquet multiplier of the cycle determines the amplitude of these terms. Our derivation implies that the second-order terms are small compared to the KS terms only for $\e\ll |\kp|$, while they cannot be neglected for moderate coupling. Generally, high-order phase reduction yields qualitative and quantitative improvement in describing the dynamics. The illustration for the former case is increased precision in determining the synchronization domain of the harmonically forced van der Pol oscillator ~\cite{mau2023c}. Examples for the latter case include an explanation of the remote synchrony in a motif of three coupled SL systems~\cite{kumar2021}, coupling-dependent effects in chimera dynamics shown in Fig.~\ref{Fig:MainResults}, and delay-induced synchronization~\cite{bick2024}.

In addition, we emphasize that the demonstrated effect
of high-order terms on the chimera's shape 
applies to a much broader range of models and phenomena.
Indeed, the standard theory says that for general 
limit-cycle oscillators and arbitrary network topology, the first-order 
approximation (the Kuramoto-Daido form) reads
$\dot{\phi}_i = \w_i + \e \mR_i(\Vec{\phi}-\phi_i \Vec{e})$,
where $\mR_i(\Vec{\phi}-\phi_i \Vec{e})$ are $2\pi$-periodic functions
which generally contain high harmonics, not only the first one.
Nonetheless, in the case of identical oscillators $\w_i=\w$,
we again eliminate the coupling parameter $\e$
by moving to the co-rotating frame and rescaling time.
Therefore, a Kuramoto-Daido first-order phase model cannot, in principle,
describe the effect of the coupling strength on the shape of any coherence-incoherence pattern in a network.
Thus, incorporating the second-order terms $\sim \e^2$ becomes crucial.

Finally, we mention that general results~\cite{mau2023c} provide a route to computing third- and higher-order terms, leading to quadruplet and so on terms, though this task is tedious and highly time-consuming. Another conclusion is that multi-body interaction appears for networks of arbitrary limit-cycle oscillators, though the closed analytical derivation of the second-order terms becomes unfeasible and must be partially done numerically, cf.~\cite{mau2023c}.
We emphasize that there exists an entirely different scenario resulting in multi-body interaction terms due to nonlinear coupling~\cite{Komarov-Pikovsky-11,*Komarov-Pikovsky-15a,*ashwin2016,*Pikovsky-Rosenblum-22} which we do not address here.

We believe our results contribute
to a better understanding of the nature
of phase models with multi-body interactions.
In particular, the results suggest physically meaningful types of such models, avoiding speculative guesswork in choosing their structure.
  
E.T.K.M. acknowledges financial support from Deutsche Forschungsgemeinschaft (DFG, German Research Foundation), Project-ID 424778381 – TRR 295.
The work of O.E.O. was supported by the Deutsche Forschungsgemeinschaft under Grant No. OM 99/2-2.
We thank A. Pikovsky for helpful discussions.



%

\end{document}